# Spontaneous valley polarization in 2D organometallic lattice

Rui Peng, Zhonglin He, Qian Wu, Ying Dai[*], Baibiao Huang, Yandong Ma[*]

School of Physics, State Key Laboratory of Crystal Materials, Shandong University, Shandanan Street 27, Jinan 250100, China

*Corresponding author: daiy60@sdu.edu.cn (Y.D.); yandong.ma@sdu.edu.cn (Y.M.)

2D ferrovalley materials that exhibit spontaneous valley polarization are both fundamentally intriguing and practically appealing to be used in valleytronic devices. Usually, the research on 2D ferrovalley materials is mainly focused on inorganic systems, severely suffering from in-plane magnetization. Here, we alternatively show by **k·p** model analysis and high-throughput first-principles calculations that ideal spontaneous valley polarization is present in 2D organometallic lattice. We explore the design principle for organic 2D ferrovalley materials composed of (quasi-)planer molecules and transition-metal atoms in hexagonal lattice, and identify twelve promising candidates. These systems have a ferromagnetic or antiferromagnetic semiconducting state, and importantly they exhibit robust out-of-plane magnetization. The interplay between spin and valley, together with strong spin-orbit coupling of transition-metal atoms, guarantee the spontaneous valley polarization in these systems, facilitating the anomalous valley Hall effect. Our findings significantly broaden the scientific and technological impact of ferrovalley physics.

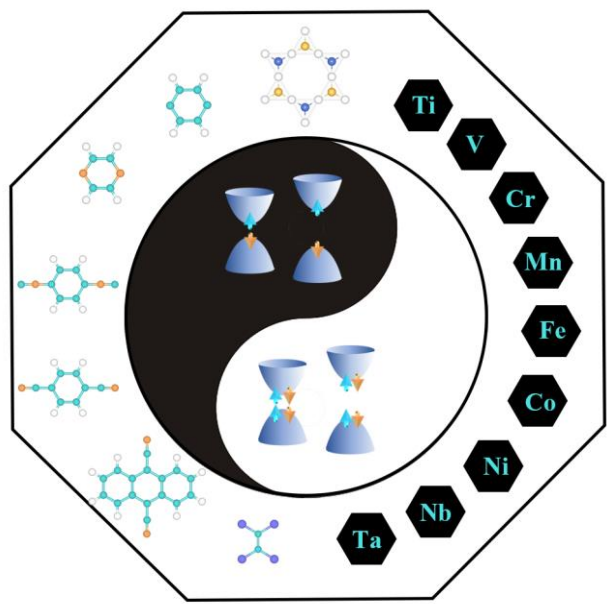

# 1. INTRODUCTION

Valley, characterizing the energy extreme of conduction or valance band, is a new degree of freedom of carriers in addition to charge and spin [1-4]. The valley index, which is robust against impurity and phonon scatterings due to its large separation in momentum space, can constitute the binary logic states in solids, yielding possible applications for information processing [1-4]. To make use of valley index as information carrier, the crucial step is to control the number of carriers in these valleys, thereby generating valley polarization. Among numerous strategies [5-12], coupling valley to intrinsic magnetic order is the most promising and attractive, where the valley polarization occurs spontaneously [13-18]. The two-dimensional (2D) systems hosting spontaneous valley polarization are referred to as 2D ferrovalley materials [14]. Depending on the type of magnetic coupling, 2D ferrovalley materials exhibit different valley-contrasting physics. Specifically, when valley couples with intrinsic ferromagnetic (FM) order, the valley-dependent optical selection rule and Berry curvature are preserved [14-18], while for intrinsic antiferromagnetic (AFM) order, the product of spin and valley indices emergences as a new electronic degree of freedom, leading to spin-valley-dependent optical selection rule and Berry curvature [13].

Though highly valuable, until now, only a few 2D ferrovalley materials have been proposed. Examples include $MnPX_3$ [13], $VSe_2$ [14], VSSe [15], $Nb_3I_8$ [16], $LaBr_2$ [17] and $AgVP_2Se_6$ [18]. Even for these few existing systems, most of them suffer from in-plane magnetization. Physically, the in-plane magnetization forbids the appearance of valley polarization, and additional tuning magnetization orientation from in-plane to out-of-plane is required [16]. This poses an outstanding challenge for the field of valleytronics. Candidate ferrovalley systems with intrinsic out-of-plane magnetization remain highly desired both from fundamental perspective and for potential use in valleytronics. Meanwhile, it is fascinating to note that all these existing 2D ferrovalley materials are confined to inorganic systems. On the other hand, many inorganic materials and devices have later found their way to organic counterparts, such as topological insulators [19-22], solar cells [23], superconductors [24], and light-emitting diodes [25]. Therefore, an inspiring question is whether 2D organic ferrovalley material, especially favoring out-of-plane magnetization naturally, exists.

In this work, based on $\mathbf{k}\cdot\mathbf{p}$ model analysis and high-throughput first-principles calculations, we present a scheme for achieving 2D organic ferrovalley materials in organometallic lattice, and obtain twelve candidates. Designed by assembling (quasi)planer molecules and transition-metal atoms with strong spin-orbit coupling (SOC) in hexagonal lattice, these candidate systems are shown to exhibit spontaneous spin polarization with out-of-plane magnetization. The coexistence of exchange interaction and strong SOC effect generates the spontaneous valley polarization. The valley-contrasting physics in these systems are further investigated, and intriguing phenomena like anomalous valley Hall effect is demonstrated. Our work not only tackles an outstanding challenge, but also opens a significant new direction for ferrovalley research.

## II. METHODS

Our first-principles calculations are performed based on density functional theory (DFT) [26] implemented in the Vienna ab initio simulation package (VASP) [27]. The exchange-correlation interaction is treated by the generalized gradient approximation (GGA) in form of Perdew-Burke-Ernzerhof (PBE) functional. [28] The cutoff energy and electronic iteration convergence criterion are set to 450 eV and $1 \times 10^{-5}$ eV, respectively. Structures, including lattice constants and atomic positions, are relaxed until the force on each atom is less than 0.01 eV/Å. A Monkhorst–Pack (MP) grid of $5 \times 5 \times 1$ is used to sample the Brillouin zone for organometallic frameworks with benzene, pyrazine and TTF, and a MP grid of $3 \times 3 \times 1$ is used to sample the Brillouin zone for organometallic frameworks with PDI, DCB and DCA. [29] To avoid the interaction between adjacent layers, a vacuum space of 20 Å is applied. For treating the strong correlation effect, we adopt PBE + U method with a moderately effective $U_{eff} = (U − J) = 2.0$ eV for transition-metal atoms. [30] SOC is incorporated in electronic structure calculations. For calculating Berry curvature and anomalous Hall conductivity, the maximally localized Wannier functions (MLWFs) implemented in WANNIER90 package are employed. [31]

## III. RESULTS

The properties of a crystalline material are correlated to both its lattice geometry and chemical composition. A typical example highlighting the importance of the lattice is the Kagome structure, where two characteristic Dirac bands emerge around the Fermi level and form a Dirac cone at the K point [32]. By breaking inversion symmetry of the Kagome lattice, a band gap opens at the Dirac cone, and the introduction of strong SOC and coulomb interaction will further enlarge the band gap, offering possibilities to form valley characters. [33,34] When additionally admitting the exchange interaction, the interplay between spin and valley might inspire the exotic ferrovalley physics. Keeping this principle in mind, we therefore adopt the Kagome-like lattice for assembling the (quasi-)planer organic molecules (OM) and transition-metal (TM) atoms. As schematically illustrated in **Fig. 1(a)**, the organometallic framework has a hexagonal lattice with a formula of $TM_2(OM)_3$, and contains two TM atoms and three (quasi)planer OM in each unit cell. To break the inversion symmetry of $TM_2(OM)_3$, there are two avenues: one is the chemical composition and the other is the exchange interaction. The former can be controlled by hetero TM atoms (referred to as TM1TM2-OM, TM1≠TM2), while the latter can be achieved by stablishing AFM state (referred to as TM-OM).

A four-band **k·p** model is employed to describe the energy states near the gapped Dirac point of our 2D frameworks. Note that the gapped Dirac cone mainly consists from the $p_z$ orbital of OM

according to the lattice symmetry [32]. With $C_3$ symmetry, the basis functions are chosen as:

$$|+1\rangle = \frac{1}{\sqrt{3}}(p_1 + \varepsilon p_2 + \varepsilon^* p_3),$$
$$|-1\rangle = \frac{1}{\sqrt{3}}(p_1 + \varepsilon^* p_2 + \varepsilon p_3),$$
$$|p_i\rangle = |OM_i, p_z\rangle,$$

where $\varepsilon = \exp\left(\frac{2i\tau\pi}{3}\right)$, and $\tau = \pm 1$ represent $\pm K$ points. The $d$ orbitals of TM atoms are neglected in this model. We then construct the effective Hamiltonian of TM1TM2-OM with considering SOC and exchange interaction as follows:

$$H^{eff} = v_F s_0(\tau_z \sigma_x k_x + \tau_0 \sigma_y k_y) + m s_0 \tau_0 \sigma_z + \lambda s_z \tau_z \left(\frac{\beta\sigma_z + 1}{2}\right) - \delta s_z \tau_0 \left(\frac{\alpha\sigma_z + 1}{2}\right)$$

Here, $v_F$ is the massless Fermi velocity, $\vec{k} = \vec{K} - \vec{K}_\pm$ is reciprocal vector, m is a symmetry-breaking perturbation admitting a band gap $\Delta = 2m$ for both spins, and $\lambda, \beta, \delta, \alpha$ characterize the spin splitting caused by SOC and exchange interaction. $s_i$, $\tau_i$ and $\sigma_i$ (i = x, y, z, 0) are Pauli matrices for spin, valley and isospin degrees of freedom, respectively. By diagonalizing the above Hamiltonian, we obtain the energy spectra:

$$E_{s,\tau,\sigma} = \frac{1}{2}(-s\delta + s\tau\lambda + \sigma\sqrt{(2m - s(\alpha\delta - \tau\beta\lambda))^2 + 4k_y^2 v_F^2 + 4k_x^2 v_F^2}),$$

where $s, \tau, \sigma = \pm 1$ represent spin, valley and band edge indices, respectively. The schematic bands around the valleys are shown in **Fig. 1(b)**. For the conduction band, $\sigma = 1$, and

$$E_{s,\tau,1} = \frac{1}{2}(-s\delta + s\tau\lambda + \sqrt{(2m - s(\alpha\delta - \tau\beta\lambda))^2 + 4k_y^2 v_F^2 + 4k_x^2 v_F^2}).$$

The energies at $\pm K$ points of conduction band can be expressed as:

$$E_c^+ = m - \frac{1+\alpha}{2}\delta + \frac{1+\beta}{2}\lambda,$$
$$E_c^- = m - \frac{1+\alpha}{2}\delta - \frac{1+\beta}{2}\lambda.$$

We can see that a valley polarization of $(1 + \beta)\lambda$ is induced in the conduction band. Similarly, we can obtain a valley polarization of $-(1 - \beta)\lambda$ in the valence band. It should be noted that as TM1≠TM2, this result is applicable for TM1TM2-OM under both FM and ferrimagnetic (FiM) states.

While for TM-OM under the AFM state, isospin is taken into consideration. The effective Hamiltonian with considering SOC is written as:

$$H^{eff} = v_F s_0(\tau_z \sigma_x k_x + \tau_0 \sigma_y k_y) + m s_0 \tau_0 \sigma_z + \delta s_z \tau_z \left(\frac{\gamma\sigma_z + 1}{2}\right),$$

where $\delta$ and $\gamma$ are renormalization of the valley gaps. The energy spectra is described as:

$$E_{\tau,\sigma} = \frac{1}{2}(\sigma\tau\delta + \sigma\sqrt{(2m + \tau\gamma\delta)^2 + 4k_y^2 v_F^2 + 4k_x^2 v_F^2}).$$

**Fig. 1(b)** presents the corresponding schematic bands around the valleys. The energies at $\pm K$ points of the conduction band can be expressed as:

$$E_c^+ = m + \frac{1+\gamma}{2}\delta$$

$$E_c^- = m - \frac{1+\gamma}{2}\delta$$

And a valley polarization of $(1+\gamma)\delta$ is induced in the conduction band of TM-OM. While for the valence band, the valley polarization is $-(1-\gamma)\delta$. Accordingly, the realization of exotic spontaneous valley polarization is feasible in both TM1TM2-OM and TM-OM in principle.

After establishing the general rule to hunt for spontaneous valley polarization in 2D organometallic lattice, we discuss its realization in real materials. To introduce exchange interaction and strong SOC effect into the lattice, we choose heavy metal atoms (i.e., Ti, V, Cr, Mn, Fe, Co, Ni, Nb and Ta) as TM candidates. And to ensure the 2D hexagonal lattice, OM with a (quasi-)planer structure, such as benzene, pyrazine, PDI, DCB, DCA and TTF [as shown in **Fig. 1(c)**], are adopted. [35-40] For TM1TM2-OM, the hetero TM atoms are considered within adjacent ones, as their atomic radius are close to each other, thus benefiting for structure stability. This results in 96 candidate systems, including 42 TM1TM2-OM and 54 TM-OM. **Fig. S1** presents the six types of crystal structures with different OM. It can be seen that due to steric repulsion, benzene, pyrazine and TTF in the lattices prefer to rotate on the TM-TM axis, resulting in a quasi-planar geometry, while for PDI, DCB and DCA, the structures favor a planar geometry without out-of-plane distortion.

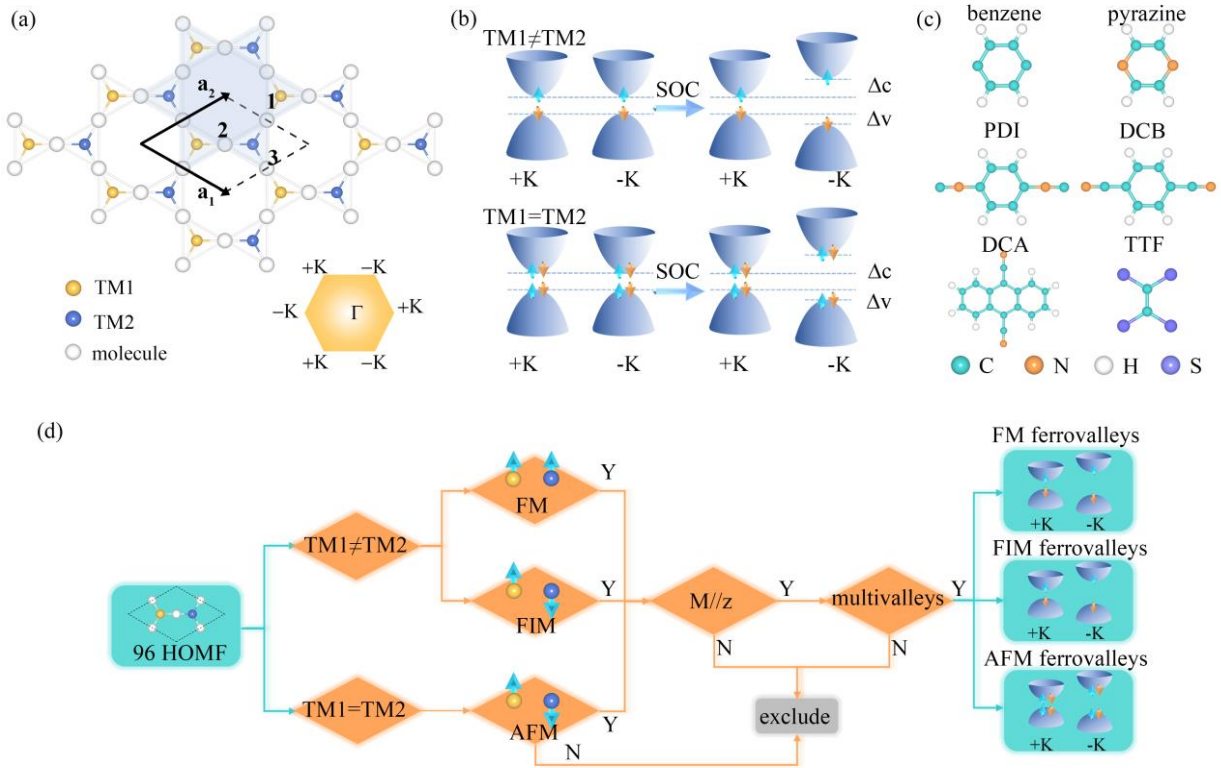

**Fig. 1** (a) Schematic structure of the organometallic framework with a formula of TM$_2$(OM)$_3$, and 2D Brillouin zone with marking the Γ and ±K points. (b) Schematic band structures near the ±K valleys without (left) and with (right) SOC. Δc and Δv in (b) represent the valley polarizations at conduction and valence bands, respectively. (c) Organic molecules chosen for constructing organometallic frameworks. (d) The algorithm for screening 2D organic ferrovalley materials. The up and down arrows in (b, d) represent up- and down-spin, respectively.

The screening process is schematically shown in **Fig. 1(d)**. We first investigate the magnetic ground states of these 96 candidates by comparing the total energies of AFM/FiM, FM and nonmagnetic (NM) states. For TM-OM, FM systems are excluded as they harbor inversion symmetry. Note that out-of-plane magnetization (M//z) is a key ingredient for realizing valley polarization, and most of the existing 2D inorganic ferrovalley materials suffer from in-plane magnetization [13-18]. We then investigate the magnetocrystalline anisotropy energy (MAE) for estimating the easy axis. The obtained magnetic ground state and magnetization easy axis for these candidate systems are summarized in **Table S1** and **S2**. Interestingly, 27 candidates, including 15 TM1TM2-OM and 12 TM-OM, are screened out following the above two criterions. Physically, the valley feature should be significant, as 'flat' valley cannot stablish the carriers, and the valley should not be submerged in the trivial bands. To satisfy this criterion, we further exclude 15 systems, and finally achieve 12 organic ferrovalley materials, including 8 TM1TM2-OM and 4 TM-OM. It is worthy emphasizing that the MAE of NbTa-benzene, NbTa-pyrazine, NbTa-DCB and Ni-DCA reach up to 17.23, 11.10, 6.00 and 6.88 meV, respectively, which is significantly larger than those of 2D inorganic ferrovalley materials [17].

**Fig. 2** shows the band structures of these 8 TM1TM2-OM with considering SOC. Arising from the large spin polarization induced by TM atoms, the spin-up and spin-down bands are significantly split. And intriguingly, as shown in **Fig. 2**, the spin-up and spin-down bands behave like being 'mutually exclusive' in the low-energy area, which is in favor of utilizing the possible valley feature. The two pairs of valleys evaluated from the deformed Dirac cones around the Fermi level for these systems are highlighted in orange shadows in **Fig. 2**. It can be seen that for NbTa-benzene, NbTa-pyrazine and MnFe-TTF, the conduction and valance band edges from one spin channel form the valleys. While for CrMn-pyrazine and NbTa-DCB, the resultant valleys shift above the Fermi level, and for TiV-pyrazine, VCr-pyrazine and CrMn-DCA, they have valleys located below the Fermi level. The valleys for these systems are all located at the ±K points. For convenience of our discussion, we refer to these two valley-related bands as $\Pi_a$ and $\Pi_b$ bands, respectively.

To quantitively characterize the spontaneous valley polarization, we define $\Delta$ as the absolute value of energy difference between the +K and –K valleys. As listed in **Table. S3**, the spontaneous valley polarization $\Delta$ of these 8 TM1TM2-OM vary from 0.03 to 278.42 meV. This significant difference in spontaneous valley polarization $\Delta$ can be attributed to the fact that except for OM-$p_z$ orbitals, TM-$d$ orbitals also contribute to the states around the valley due to the orbital coupling between TM and

OM. Taking NbTa-benzene and CrMn-pyrazine as examples, there are relatively more contributions from the TM-$d$ orbitals for the former one. As TM-$d$ orbitals exhibit stronger SOC strength with respect to OM-$p_z$ orbitals, larger valley polarization is obtained in NbTa-benzene as compared with that of CrMn-pyrazine. Particularly, among these systems, the spontaneous valley polarization in the $\Pi_a$ band of NbTa-benzene is as large as 278.42 meV, much larger than those of all the previously reported 2D inorganic systems [13-18]. Such giant spontaneous valley polarization is equivalent to the case in which a valley degenerate system is under a magnetic field of 1392 T, and is robust against the external perturbations. This, combined with the stable out-of-plane magnetization, indicate that these systems are promising organic ferrovalley materials. Another interesting point we wish to address is that in addition to these two pairs of valleys, there are several other valleys in the band structure, as shown in **Fig. 2**. And intriguingly for some systems, these valleys also exhibit significantly spontaneous valley polarization. For example, the valleys at the second-highest valence band of NbTa-benzene present a spontaneous valley polarization of 29 meV, and the valleys at the second-lowest conduction band of NbTa-pyrazine show a spontaneous valley polarization of 120 meV. Such multivalley characteristic, especially with large spontaneous valley polarization, is seldom reported in inorganic materials [13-18].

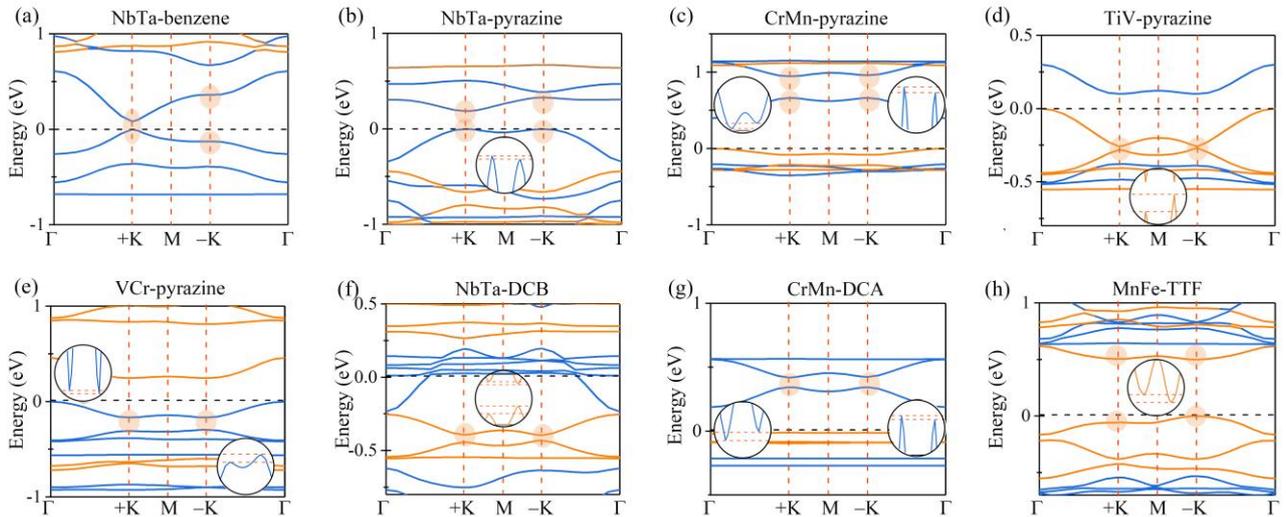

**Fig. 2** Band structures of (a) NbTa-benzene, (b) NbTa-pyrazine, (c) CrMn-pyrazine, (d) TiV-pyrazine, (e) VCr-pyrazine, (f) NbTa-DCB, (g) CrMn-DCA, and (h) MnFe-TTF. The two pairs of valleys evaluated from the deformed Dirac cones around the Fermi level for these systems are highlighted in orange shadows. Fermi level is set to 0 eV.

The band structures of the 4 TM-OM with considering SOC are presented in **Fig. 3**. Because of the AFM coupling, the spin-up and spin-down channels of TM-OM are degenerate in energy. For V-pyrazine and Cr-PDI, the two pairs of valleys evaluated from the deformed Dirac cones are located around the Fermi level [**Fig. 3(b)** and **(c)**], while for Ni-pyrazine and Ni-DCA, the valleys are located above the Fermi level [**Fig. 3(a)** and **(d)**]. The spontaneous valley polarization of these systems are summarized in **Table. S4**, from which we can see that all these 4 systems exhibit sizeable spontaneous valley polarization. Especially for Ni-pyrazine, the spontaneous valley polarization in the $\Pi_a$ band reaches up to 94.48 meV. Therefore, these 4 TM-OM are tantalizing organic ferrovalley materials as well. Moreover, by staring at the band structure of Ni-DCA shown in **Fig. 3(d)**, we can see that its valley features are special as compared with those of the other three systems. That is, the valleys from the $\Pi_a$ and $\Pi_b$ bands at the –K point are almost degenerate in energy, showing an energy gap of only 1.04 meV, while it is 38.06 meV for the case at the +K point. Therefore, Ni-DCA can be considered as a half-valley metal, wherein the conduction carriers are 100% valley polarized, when shifting the Fermi level to between the $\Pi_a$ and $\Pi_b$ bands. To our knowledge, the half-valley metal has only been virtually achieved in inorganic monolayer H-FeCl$_2$ by tuning the electron correlation effect so far [41]. Different from this hypothetical case, the half-valley metallic feature in Ni-DCA is intrinsic, although there is a negligible energy gap.

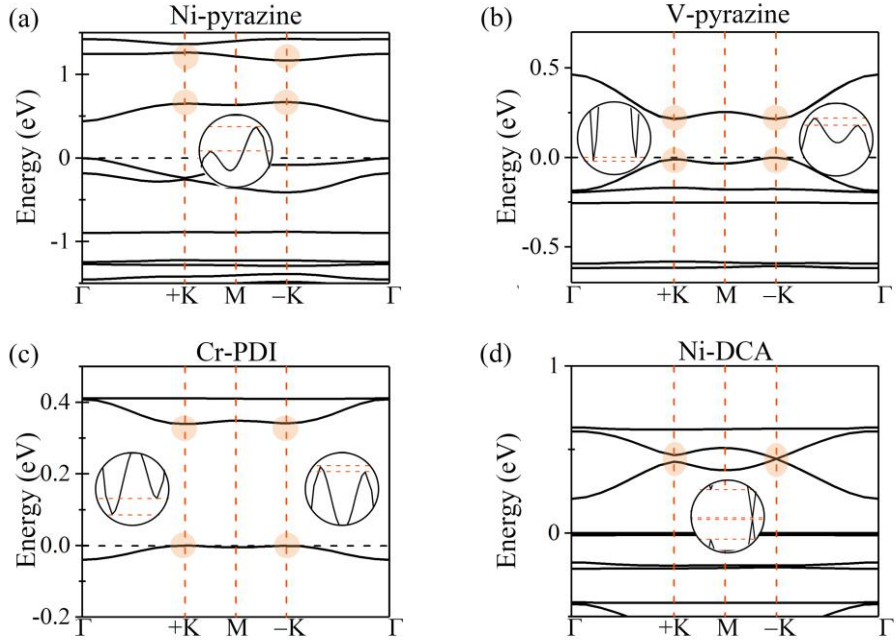

**Fig. 3** Band structures of (a) Ni-benzene, (b) V-pyrazine, (c) Cr-PDI, and (d) Ni-DCA. The two pairs of valleys evaluated from the deformed Dirac cones around the Fermi level for these systems are highlighted in orange shadows. Fermi level is set to 0 eV.

Then we discuss the valley-contrasting physics in these 2D organic ferrovalley materials. Due to the broken inversion symmetry, the low-energy quasiparticle states of these 2D organic ferrovalley materials at +K and –K valleys are assigned a valley index $\tau = \pm 1$. The orbital magnetic moments of the +K/–K valleys in FM (FiM) and AFM systems can be expressed as

$$\mathcal{M} = \tau \mu_B^*$$

and

$$\mathcal{M} = 2s\tau\mu_B^*,$$

respectively [13]. Here, $\mu_B$ is the effective Bohr magneton, and s is the spin index. Based on the orbital magnetic moment $\mathcal{M}$, the Berry curvatures of these 2D organic ferrovalley materials at the ±K points can be expressed as:

$$\Omega(\pm K) = \hbar\mathcal{M}(\pm K)/e\varepsilon(\pm K),$$

where e and $\varepsilon(\pm K)$ are the electronic charge and band energy at the valleys, respectively [13]. Obviously, for the 8 TM1TM2-OM, the Berry curvature $\Omega(\pm K)$ is valley-dependent, while for the 4 TM-OM, it is dependent on the product of valley and spin indices.

To get more insight into the valley-contrasting physics in these 8 FM (FiM) systems, we take NbTa-pyrazine as a representation to calculate the Berry curvature $\Omega$ on the basis of the Kudo formula, which can be written as:[42]

$$\Omega(k) = -\sum_n \sum_{n \neq n'} f_n \frac{2Im\langle\psi_{nk}|v_x|\psi_{n'k}\rangle\langle\psi_{n'k}|v_y|\psi_{nk}\rangle}{(E_n-E_{n'})^2},$$

Here, $f_n$ is the Fermi-Dirac distribution function, $\psi_{nk}$ is the Bloch wave function with eigenvalue $E_n$, and $v_{x/y}$ is the velocity operator along x/y direction. The Berry curvature $\Omega$ over the entire 2D Brillion zone and along the high-symmetry points is plotted in **Fig. 4(a, b)**. Clearly, the Berry curvatures $\Omega$ at +K and –K valleys have opposite signs, and their absolute values are different due to SOC. While for the other area of the Brillion zone, the Berry curvature is almost zero. When shifting the Fermi level between the +K and –K valleys of the $\Pi_a$ band, the spin-down electrons from +K valley will acquire an anomalous velocity [i.e., v~E × $\Omega(k)$] under an in-plane electric field and accumulate at the bottom edge of the sample; see **Fig. 4(c)**. And when shifting the Fermi level between the +K and –K valleys of the $\Pi_b$ band, the spin-down holes from +K valley will acquire an opposite anomalous velocity with respect to the former case under an in-plane electric field and accumulate at the top edge of the sample. In both cases, a net charge/spin/valley current is generated, realizing the exotic anomalous valley Hall effect. In addition to doping carriers, linearly polarized light can also be adopted to achieve the anomalous valley Hall effect in NbTa-pyrazine. Upon exciting the spin-up electrons and spin-down holes at the +K valley via linearly polarized light, the in-plane electric field can drive these electrons and holes to opposite boundaries of the sample as they are subjected to different Berry curvatures $\Omega$, forming the anomalous charge/spin/valley Hall effect as well.

The valley-contrasting physics in these 4 AFM systems are different from these 8 FM (FiM) systems. Taking V-pyrazine as an example, we calculate its Berry curvatures and present them in **Fig. 4(a, b)**. We can see that for +K (–K) valleys, the Berry curvatures are opposite for spin-up and spin-down states. When shifting the Fermi level between the +K and –K valleys of the $\Pi_a$ band, the spin-down and spin-up electrons from +K valley will acquire opposite anomalous velocities under an in-plane electric field and accumulate at the bottom and top edges of the sample, respectively. Therefore, a net spin current is generated, contributing to the spin Hall effect, but not the anomalous valley Hall effect; see **Fig. 4(d)**. To realize anomalous valley Hall effect in V-pyrazine, circularly polarized light is needed. Owing to the preserved joint operation $\hat{T}\hat{P}$, where $\hat{T}$ and $\hat{P}$ represent time reversal and inversion symmetry, we have the following results at +K valley:

$$\langle B, +1, \uparrow |\sigma_+|A, -1, \uparrow\rangle = \langle B, +1, \uparrow |C_3^{-1}C_3\sigma_+C_3^{-1}C_3|A, -1, \uparrow\rangle$$
$$= exp(-i2\pi/3)\langle B, +1, \uparrow |\sigma_+|A, -1, \uparrow\rangle$$

$$\langle A, -1, \downarrow |\sigma_+|B, +1, \downarrow\rangle = \langle A, -1, \downarrow |C_3^{-1}C_3\sigma_+C_3^{-1}C_3|B, +1, \downarrow\rangle = \langle A, -1, \downarrow |\sigma_+|B, +1, \downarrow\rangle$$

$$\langle B, +1, \uparrow |\sigma_-|A, -1, \uparrow\rangle = \langle B, +1, \uparrow |C_3^{-1}C_3\sigma_-C_3^{-1}C_3|A, -1, \uparrow\rangle = \langle B, +1, \uparrow |\sigma_-|A, -1, \uparrow\rangle$$

$$\langle A, -1, \downarrow |\sigma_-|B, +1, \downarrow\rangle = \langle A, -1, \downarrow |C_3^{-1}C_3\sigma_-C_3^{-1}C_3|B, +1, \downarrow\rangle$$
$$= exp(i2\pi/3)\langle A, -1, \downarrow |\sigma_-|B, +1, \downarrow\rangle,$$

where we have used $C_3\sigma_+C_3^{-1} = \exp(i2\pi/3)\sigma_+, C_3\sigma_-C_3^{-1} = \exp(-i2\pi/3)\sigma_-$. A and B represent two sublattices, ±1 represent two basic functions, ↓ and ↑ represent spin down and spin up states. Accordingly, under the left circularly polarized light $\sigma_+$ with appropriate frequency, only the spin-down electrons and spin-up holes at +K valley will be generated and acquire opposite anomalous velocities under an in-plane electric field. As a result, they would accumulate at opposite edges of the sample, as shown in **Fig. 4(d)**. While under the right circularly polarized light $\sigma_-$ with the same frequency, only the spin-up electrons and spin-down holes at +K valley will be generated and accumulate at opposite edges of the sample. In these two cases, the net charge/spin/valley current is achieved, yielding the anomalous charge/spin/valley Hall effect.

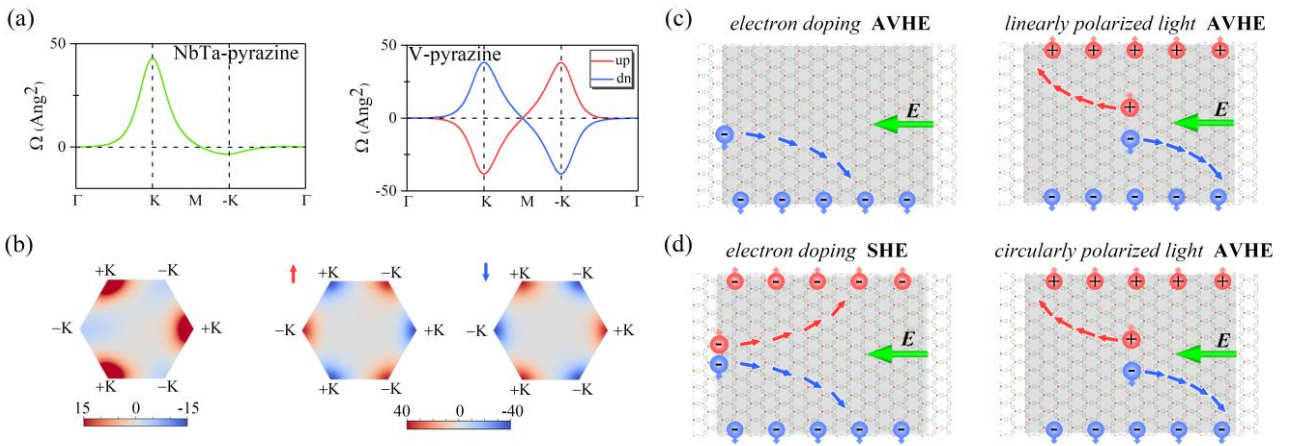

**Fig. 4** (a) Berry curvatures of NbTa-pyrazine, and the spin-up and spin-down valence bands of V-pyrazine along the high-symmetry points. (b) Berry curvatures of NbTa-pyrazine, and the spin-up and spin-down valence bands of V-pyrazine over the entire 2D Brillouin zone. (c) Diagrams of the anomalous valley Hall effect in NbTa-pyrazine under electron doping and linearly light irradiation. (d) Diagrams of the spin Hall effect under electron doping and the anomalous valley Hall effect under left circularly light irradiation in V-pyrazine. The "+" ("–") symbols represent electrons (holes), and the red and blue arrows represent spin-up and spin-down states, respectively.

## IV. CONCLUSION

To summarize, we for the first time propose a design principle for achieving spontaneous valley polarization in organometallic lattice via **k·p** model analysis. And through high-throughput first-principles calculations, we further identify its realization in 12 real 2D organic materials, including 8 TM1TM2-OM and 4 TM-OM. Importantly, as a new ferrovalley family, these systems exhibit out-of-plane magnetization naturally. In addition, we predict that such systems could demonstrate many intriguing phenomena, for example, multivalley characteristic, extremely large valley polarization and anomalous valley Hall effect. The underlying physics are discussed in detail. Our work opens a new direction for valleytronic research and would promote technological innovation.

## ACKNOWLEDGEMENT

This work is supported by the National Natural Science Foundation of China (No. 11804190, 12074217), Shandong Provincial Natural Science Foundation (Nos. ZR2019QA011 and ZR2019MEM013), Shandong Provincial Key Research and Development Program (Major Scientific and Technological Innovation Project) (No. 2019JZZY010302), Shandong Provincial Key Research and Development Program (No. 2019RKE27004), Shandong Provincial Science Foundation for Excellent Young Scholars (No. ZR2020YQ04), Qilu Young Scholar Program of Shandong University, and Taishan Scholar Program of Shandong Province.

## REFERENCE

[1] D. Xiao, W. Yao, and Q. Niu, Valley-contrasting physics in graphene: magnetic moment and topological transport, Phys. Rev. Lett. 99, 236809 (2007).
[2] W. Yao, D. Xiao, and Q. Niu, Valley-dependent optoelectronics from inversion symmetry breaking, Phys. Rev. B 77, 235406 (2008).
[3] D. Xiao, G.-B. Liu, W. Feng, X. Xu, and W. Yao, Coupled spin and valley physics in monolayers of $MoS_2$ and other Group-VI dichalcogenides, Phys. Rev. Lett. 108, 196802 (2012).
[4] X. Xu, W. Yao, D. Xiao, and T. F. Heinz, Spin and pseudospins in layered transition metal dichalcogenides, Nat. Phys. 10, 343 (2014).
[5] K. F. Mak, K. He, J. Shan, and T. F. Heinz, Control of valley polarization in monolayer $MoS_2$ by optical helicity, Nat. Nanotech. 7, 494 (2012).
[6] H. Zeng, J. Dai, W. Yao, D. Xiao, and X. Cui, Valley polarization in $MoS_2$ monolayers by optical pumping, Nat. Nanotech. 7, 490 (2012).
[7] X. Xu, Y. Ma, T. Zhang, C. Lei, B. Huang, and Y. Dai, Nonmetal-atom-doping-induced valley polarization in single-layer $Tl_2O$, J. Phys. Chem. Lett. 10, 4535 (2019).
[8] R. Peng, Y. Ma, S. Zhang, B. Huang, and Y. Dai, Valley polarization in Janus single-layer MoSSe via magnetic doping, J. Phys. Chem. Lett. 9, 3612 (2018).
[9] Z. Zhang, X. Ni, H. Huang, L. Hu, and F. Liu, Valley splitting in the van der Waals heterostructure $WSe_2/CrI_3$: The role of atom superposition, Phys. Rev. B 99, 115441 (2019).
[10] M. K. Mohanta, and A. D. Sarkar, Coupled spin and valley polarization in monolayer $HfN_2$ and valley-contrasting physics at the $HfN_2$-$WSe_2$ interface, Phys. Rev. B 102, 125414 (2020).

[11] G. Aivazian, Z. Gong, A.M. Jones, R.-L. Chu, J. Yan, D. G. Mandrus, C. Zhang, D. Cobden, W. Yao, and X. Xu, Magnetic control of valley pseudospin in monolayer $WSe_2$, Nat. Phys. 11, 148 (2015).

[12] X.-X. Zhang, T. Cao, Z. Lu, Y.-C. Lin, F. Zhang, Y. Wang, Z. Li, J. C. Hone, J. A. Robinson, D. Smirnov, S. G. Louie, and T. F. Heinz, Magnetic brightening and control of dark excitons in monolayer $WSe_2$, Nat. Nanotech. 12, 883 (2017).

[13] X. Li, T. Cao, Q. Niu, J. Shi, and J. Feng, Coupling the valley degree of freedom to antiferromagnetic order, Proc. Natl. Acad. Sci. U. S. A. 110, 3738 (2013).

[14] W.-Y. Tong, S.-J. Gong, X. Wan, and C.-G. Duan, Concepts of ferrovalley material and anomalous valley Hall effect, Nat. Commun. 7, 13612 (2016).

[15] C. Zhang, Y. Nie, S. Sanvito, and A. Du, First-principles prediction of a room-temperature ferromagnetic Janus VSSe monolayer with piezoelectricity, ferroelasticity, and large valley polarization, Nano Lett. 19, 1366 (2019).

[16] R. Peng, Y. Ma, X. Xu, Z. He, B. Huang, and Y. Dai, Intrinsic anomalous valley Hall effect in single-layer $Nb_3I_8$, Phys. Rev. B 102, 035412 (2020).

[17] P. Zhao, Y. Ma, C. Lei, H. Wang, B. Huang, and Y. Dai, Single-layer $LaBr_2$: Two-dimensional valleytronic semiconductor with spontaneous spin and valley polarizations, Appl. Phys. Lett. 115, 261605 (2019).

[18] Z. Song, X. Sun, J. Zheng, F. Pan, Y. Hou, M.-H. Yung, J. Yang, and J. Lu, Spontaneous valley splitting and valley pseudospin field effect transistors of monolayer $VAgP_2Se_6$, Nanoscale 10, 13986 (2018).

[19] Z. Wang, Z. Liu, and F. Liu, Organic topological insulators in organometallic lattices, Nat. Commun. 4, 1471 (2013).

[20] Z. Wang, Z. Liu, and F. Liu, Quantum anomalous Hall effect in 2D organic topological insulators, Phys. Rev. Lett. 110, 196801 (2013).

[21] Z. Liu, Z. Wang, J. Mei, Y. Wu, and F. Liu, Flat Chern band in a two-dimensional organometallic framework, Phys. Rev. Lett. 110, 106804 (2013).

[22] X. Wu, Y. Feng, S. Li, B. Zhang, and G. Gao, 2D $Mn_2C_6Se_{12}$ and $Mn_2C_6S_6Se_6$: Intrinsic room-temperature Dirac spin gapless semiconductors and perfect spin transport properties, J. Phys. Chem. C 124, 16127 (2020).

[23] D. Kearns, and M. Calvin, Photovoltaic effect and photoconductivity in laminated organic systems, J. Chem. Phys. 29, 950 (1958).

[24] D. Jerome, A. Mazaud, M. Ribault, and K. Bechgaard, Superconductivity in a synthetic organic conductor (TMTSF)2PF6. J. Phys. Lett. (Paris) 41, L95 (1980).

[25] C. Tang, and S. A. Vanslyke, Organic electroluminescent diodes. Appl. Phys. Lett. 51, 913 (1987).

[26] W. Kohn and L. J. Sham, Self-consistent equations including exchange and correlation effects, Phys. Rev. 140, A1133 (1965).

[27] G. Kresse and J. Furthmüller, Efficient iterative schemes for ab initio total-energy calculations using a plane-wave basis set, Phys. Rev. B 54, 11169 (1996).

[28] J. P. Perdew, K. Burke, and M. Ernzerhof, Generalized gradient approximation made simple, Phys. Rev. Lett. 77, 3865 (1996).

[29] H. J. Monkhorst and J. D. Pack, Special points for Brillouin-zone integrations, Phys. Rev. B 13, 5188 (1976).

[30] L. Wang, T. Maxisch, and G. Ceder, Oxidation energies of transition metal oxides within the GGA+ U framework, Phys. Rev. B: Condens. Matter 73, 195107 (2006).

[31] A. A. Mostofi, J. R. Yates, G. Pizzi, Y.-S. Lee, I. Souza, D. Vanderbilt, and N. Marzari, An updated version of wannier90: A tool for obtaining maximally-localised Wannier functions, Comput. Phys. Commun. 185, 2309 (2014).

[32] E. Tang, J. Mei, and X. Wen, High-temperature fractional quantum Hall states, Phys. Rev. Lett. 106, 236802 (2011).

[33] M. Zhao, A. Wang, and X. Zhang, Half-metallicity of a kagome spin lattice: the case of a manganese bis-dithiolene monolayer, Nanoscale 5, 10404 (2013).

[34] Xu Guo, Zhao Liu, Bing Liu, Qunxiang Li, and Zhengfei Wang, Non-Collinear Orbital-induced Planar Quantum Anomalous Hall Effect, Nano Lett. 20, 7606 (2020).

[35] Y. Jin, Z. Chen, B. Xia, Y. Zhao, R. Wang, and H. Xu, Large-gap quantum anomalous Hall phase in hexagonal organometallic frameworks, Phys. Rev. B 98, 245127 (2018).

[36] J. P. Darby, M. Arhangelskis, A. D. Katsenis, J. M. Marrett, T. Friščić, and A. J. Morris, Ab initio prediction of metal-organic framework structures, Chem. Mater. 32, 5835 (2020).

[37] H. Sun, S. Tan, M. Feng, J. Zhao, and H. Petek, Deconstruction of the electronic properties of a topological insulator with a two-dimensional noble metal−organic honeycomb−Kagome band structure. J. Phys. Chem. C 122, 18659 (2018).

[38] C. Hsu, Z. Huang, G. M. Macam, F. Chuang, and L. Huang, Prediction of two-dimensional organic topological insulator in metal-DCB lattices, Appl. Phys. Lett. 113, 233301 (2018).

[39] L. Z. Zhang, Z. F. Wang, B. Huang, B. Cui, Z. Wang, S. X. Du, H.-J. Gao, and F. Liu, Intrinsic two-dimensional organic topological insulators in metal−dicyanoanthracene lattices, Nano Lett. 16, 2072 (2016).

[40] A. Wang, X. Zhang, Y. Feng, and M. Zhao, Chern insulator and Chern half-metal states in the two-dimensional spin-gapless semiconductor $Mn_2C_6S_{12}$, J. Phys. Chem. Lett. 8, 3770 (2017).

[41] H. Hu, W. Tong, Y. Shen, X. Wan, and C. Duan, Concepts of the half-valley-metal and quantum anomalous valley Hall effect, npj Comput. Mater. 6, 129, (2020).

[42] D. J. Thouless, M. Kohmoto, M. P. Nightingale, and M. den Nijs, Quantized Hall conductance in a two-dimensional periodic potential, Phys. Rev. Lett. 49, 405 (1982).